\documentstyle[aps,prl,epsf,twocolumn,floats]{revtex} 

\newcommand{\GE}{{\Gamma}} 
\newcommand{\se}{{y_0}} 

\begin{document} 
\draft 
\twocolumn[\csname @twocolumnfalse\endcsname 
\widetext 

\title{Universally diverging Gr\"uneisen parameter and the magnetocaloric
effect close to quantum critical points} 
\author{Lijun Zhu$^1$, Markus Garst$^2$, Achim Rosch$^2$, and Qimiao Si$^1$} 
\address{$^1$Department of Physics \& Astronomy, Rice University, Houston, 
TX 77005--1892, USA\\ 
$^2$Institut f\"ur Theorie der Kondensierten Materie, Universit\"at Karlsruhe, 
D-76128 Karlsruhe, Germany} 

\maketitle 

\begin{abstract} 
At a generic quantum critical point,
the thermal expansion
$\alpha$ is 
more singular than the specific heat $c_p$.
Consequently, the ``Gr\"uneisen ratio'', $\GE=\alpha/c_p$, diverges.
When scaling applies,
$\GE \sim T^{-1/(\nu z)}$ at the critical pressure $p=p_c$,
providing a means to measure the scaling dimension of the
most relevant operator that pressure couples to;
in the alternative limit $T\to0$ and $p \ne p_c$,
$\GE \sim \frac{1}{p-p_c}$ with a prefactor that is,
up to the molar volume, a 
simple {\it universal} combination of critical exponents.  
For a
  magnetic-field driven transition, similar relations hold for the
  magnetocaloric effect $\left.(1/T)\partial T/\partial H\right|_S$.
  Finally, we determine the corrections to scaling in a class of
  metallic quantum critical points.
\end{abstract} 
\pacs{PACS numbers: 71.10.Hf, 71.27.+a, 75.20.Hr, 71.28.+d} 
] 

\narrowtext 

{\it Introduction:} The anomalous behavior observed in an increasing
number of systems, ranging from insulating magnets and heavy fermion
compounds to cuprate superconductors, has been attributed to the
presence of quantum critical points (QCPs). These occur in systems
where a continuous quantum phase transition (QPT) at $T=0$ is induced
by tuning some control parameter like pressure $p$, doping or magnetic
field $H$. Such zero-temperature critical points can determine the
properties of materials in a wide range of temperatures.
In general, quantum critical points are more difficult to 
characterize compared to their classical counterparts.
At a classical critical point, thermodynamic quantities
typically diverge; 
the associated critical exponents historically
played a central role in our eventual understanding
of scaling and universality.
Some of these divergences, however, have to disappear at a QCP: 
there are constraints placed by the third law of thermodynamics
due to the very fact that the transition takes place at
zero temperature.
Here we
show 
that the Gr\"uneisen ratio  \cite{Grueneisen,Landau} diverges
at any QCP, in a way that provides a novel
thermodynamic means of probing quantum phase transitions.

We define the  Gr\"uneisen ratio $\Gamma$ \cite{Grueneisen,Landau,gruen}
in terms of
the molar specific heat $c_p=\frac{T}{N}
\left.\frac{\partial S}{\partial T}\right|_p$
and the
thermal expansion 
$\alpha=\frac{1}{V}  \left.\frac{\partial V}{\partial T}\right|_{p,N}=  
-\frac{1}{V}   \left.\frac{\partial S}{\partial p}\right|_{T,N}$
\begin{eqnarray}
\label{def}
\Gamma=\frac{\alpha}{c_p} = -\frac{1}{V_m T}\frac{\partial S/\partial p}{ \partial S/\partial
T} 
\end{eqnarray}
where $S$ is the entropy and $V_m=V/N$ the molar volume.
In ordinary situations, 
 pressure dependences are regular and a finite Gr\"uneisen ratio is
expected as is indeed observed in all previous measurements of this
quantity in the literature.  Such a regular dependence is typically
described by assuming that the system is dominated by a single energy
scale $E^*$ (e.g.  the Fermi energy in a metal or the Debye energy if
acoustic phonons dominate) so the molar entropy takes the form
$S/N=f(T/E^*)$.  The Gr\"uneisen ratio is then temperature independent
and given by\cite{Grueneisen,Landau,Fulde,Kambe} $\GE=\frac{1}{V_m
  E^*} \frac{\partial E^*}{\partial p}$. However, this formula already
suggests that a diverging $\Gamma$ can be expected when some energy scale $E^*$ vanishes as it happens at a QCP.

\begin{figure} 
\vbox{
\hspace{3ex}
\epsfxsize=70mm
\epsfbox{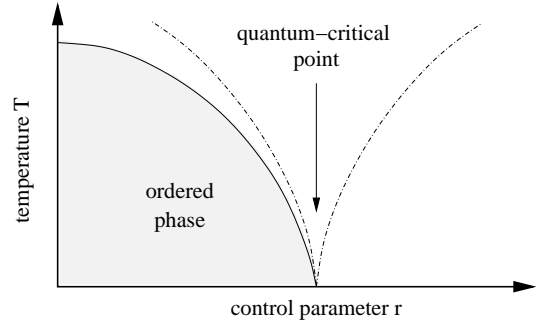}} 
\vspace{1ex} 
\caption{Schematic phase diagram with a QCP.}
\label{fig:phase-diagram} 
\end{figure} 

{\it Divergence of the Gr\"uneisen ratio at QCPs:} A quantum critical
point is reached in a singular fashion by tuning some external
parameter and, in general, this external parameter is
thermodynamically coupled to pressure.  In the low temperature limit,
the singular terms of $S$ and $T$ in Eq. (\ref{def}) cancel out
leaving $\Gamma$ to depend only on singularities associated with the
pressure $p$. As the pressure controls the QPT, such a singularity
always exists and the Gr\"uneisen ratio diverges at any QCP.  This
divergence is entirely determined by the scaling dimension of the
control parameter, which is the most relevant operator to which the
pressure couples.  As shown below, this leads to a $T$ dependence,
$\Gamma \sim 1/T^{1/\nu z}$ [see Eq.~(\ref{gr1})].  {\it In other
  words, the temperature exponent of the Gr\"uneisen ratio provides a
  direct means to measure $\nu z$ and, as a result, characterize a
  QCP.}
Put in a slightly different way,
the thermal expansion contains valuable information 
complementary to that obtained from the specific heat: 
while $c_p$
measures the response 
to $T$ (y-axis in
Fig.~\ref{fig:phase-diagram}), $\alpha$ describes the response to the
tuning parameter of the QPT, the second relevant variable at a QCP
(x-axis in Fig.~\ref{fig:phase-diagram}).
This has to be contrasted with
a classical phase transition. There, generically, only one relevant
operator exists to which both $T$ and $p$ couple.  Accordingly, $\GE$
will be constant close to a classical transition.

To observe the singular behavior of 
$\Gamma$ or the thermal expansion
the pressure has to couple sufficiently strongly 
to the critical dynamics.  This is
for example the case in heavy fermion compounds
where the 
intricate competition between magnetic interactions and the Kondo
effect can be tuned by pressure, doping or magnetic field to yield a
QPT, typically from a metallic antiferromagnet to
a metallic paramagnet.
The high sensitivity 
to pressure 
arises from the exponential dependence of the Kondo
temperature on system parameters.  Whether the transitions in these
systems conform to the Gaussian picture associated with $T=0$
spin-density wave (SDW) transitions\cite{Hertz,Millis} or are
non-Gaussian as in a locally quantum critical point\cite{Si_Nature} is
a question of great current interest.

If the control parameter of the QPT is not pressure but an 
external magnetic field $H$, the role of the Gr\"uneisen 
ratio
is played by 
the ratio of the $T$-derivative of the 
magnetization $M$ (per mole)
to
the molar specific heat, 
for either fixed pressure or fixed volume, 
\begin{eqnarray} 
\Gamma_H= 
-\frac{(\partial M/\partial T)_H}{c_H} 
= -\frac{1}{T}\frac{(\partial S/\partial H)_T} 
{(\partial S/\partial T)_H} 
= \frac{1}{T}\left.\frac{\partial T}{\partial H}\right|_S\,. 
\end{eqnarray} 
It can be determined directly from the
magnetocaloric effect
-- 
the change of temperature in
response to an adiabatic ($S = const.$) change of $H$.

In the following, we go beyond these general considerations by
carrying through a) a more detailed analysis based on the assumption
of scaling and b) a model study on the spin-density-wave (SDW) QCPs
in metallic systems. The latter is a model system that is 
above or equal to
the upper critical dimension, so corrections to scaling are important;
it is of direct interest in comparing with
experiments in heavy fermion compounds.

{\it Scaling analysis:} Close to any QCP, 
the correlation length $\xi$ diverges as a function of a control
parameter $r$, 
$\xi \sim |r|^{-\nu}$, 
where e.g. $r=(p-p_c)/p_c$ or $r=(H-H_c)/H_c$. 
Correspondingly, a typical correlation (imaginary) time, 
$\xi_\tau \sim \xi^{z} $,
diverges as the QCP is approached. The ``dynamical critical exponent''
$z$ depends on the dynamics of the order parameter and relates time
and length scales.

If one assumes that the critical behavior is governed by $\xi$ and
$\xi_\tau$ (a more careful discussion of this assumption is given
below), the critical contribution to the free energy per mole,
$F_{cr}=F-F_{\text{reg}}$, can be cast into the following standard
scaling {\it Ansatz} (using hyperscaling)
\begin{eqnarray} 
\frac {F_{cr}}{N}  &=&   
- \rho_0 \, r^{\nu (d+z)} 
\tilde{f}\!\left(\frac{T}{T_0 r^{\nu z}}\right) \nonumber\\ 
&=& 
- \rho_0 \left(\frac{T}{T_0}\right)^{(d+z)/z} 
f\!\left(\frac{r}{(T/T_0)^{1/(\nu z)}}\right) \,,
\label{scaling} 
\end{eqnarray} 
where $\rho_0$ and $T_0$ are non-universal constants, while $f(x)$ and
$\tilde{f}(x)$ are universal scaling functions. Obviously, $f(x\to
0)\approx f(0)+x f'(0)+...$ is regular as there is no phase transition
at $r=0,T>0$ (see Fig.~\ref{fig:phase-diagram}). The limit
$\tilde{f}(x\to 0)=\tilde{f}(0)+c \, x^{\se+1}$ describes the low
temperature behavior of the phases to the left or right side of the
QCP (in general different for $r>0$ and $r<0$). Note that the exponent
$\se>0$ has to be positive due to the third law of thermodynamics. It
characterizes the power-law behavior of the specific heat $c_p\sim
T^{\se}$, e.g. $\se=1$ for a Fermi liquid, $\se=2$ for a d-wave
superconductor in $d=2$, or $\se=d$ and $d/2$ for an insulating
antiferromagnet and ferromagnet, respectively.

Thermodynamical quantities are easily obtained from (\ref{scaling}).
The critical contribution $c_{cr}$ to the specific heat at $r=0$ is
given by
\begin{equation} 
c_{cr}(T, r=0) = \frac{(d+z) d}{z^2} \frac{\rho_0}{T_0} f(0) 
\left(\frac{T}{T_0}\right)^{d/z} 
\label{scaling:specificHeat} 
\end{equation} 
and for $T\to 0$, $r\neq0$ 
\begin{equation} 
c_{cr}(T\to0, r) = \frac{\rho_0 c  \se (\se+1) }{T_0} 
\left(\frac{T}{T_0}\right)^{\se} r^{\nu (d-\se z )}\;. 
\label{scaling:specificHeat2} 
\end{equation} 
Similarly, in the case of a pressure tuned QCP with $r=(p-p_c)/p_c$
the critical contribution $\alpha_{cr}$ to the thermal expansion reads
\begin{eqnarray}\label{scaling:thermalExpansion} 
\alpha_{cr}(T, r=0) &=& - 
\frac {d+z-\frac{1}{\nu}}{z} 
\frac{\rho_0 f'(0) }{T_0 p_c V_m} 
\left(\frac{T}{T_0}\right)^{(d-\frac 1{\nu})/z}
\label{alpha-T} 
\end{eqnarray} 
and 
for $r\neq 0$
\begin{eqnarray}\label{scaling:thermalExpansion2} 
\alpha_{cr}(T\to 0) &=& 
-\frac{\rho_0 (\se+1) 
c \nu(d-\se z)}{T_0 V_m}
\frac{r^{\nu (d-\se z )}}{p_c r} \left(\frac{T}{T_0}\right)^{\se} \!\!\!.\nonumber \\ 
\label{alpha-r} 
\end{eqnarray} 
The thermal expansion is more singular 
than the specific
heat leading to a Gr\"uneisen ratio
\begin{eqnarray} 
\GE_{cr}(T,r=0)=\frac{\alpha_{cr}}{c_{cr}} 
= -G_T T^{-1/(\nu z)} \,,
\label{gr1} 
\end{eqnarray} 
where the prefactor $G_T = {{(d + z - 1/\nu)z f'(0)} \over {(d+z)d\,
    f(0)}} { {T_0^{1/(\nu z)}} \over {p_c V_m}}$ contains some
non-universal parameters ($p_c$ and $T_0$).  
{\it We reach the important conclusion that the temperature exponent
of the Gr\"uneisen ratio is equal to $1/\nu z$.}

In the other limit
$T\to0$, $r \neq 0$ we obtain the universal result
\begin{eqnarray} 
\GE_{cr}(T\to 0,r)= - G_r 
\frac{1}{V_m (p-p_c)}. 
\label{gr2} 
\end{eqnarray} 
Remarkably, even the (generally unknown) scaling functions cancel out
in the amplitude $G_r$, leaving only a combination of critical
exponents and the dimensionality:
\begin{eqnarray} 
G_r = \frac{\nu (d-\se z)}{\se}\,. 
\label{Cr} 
\end{eqnarray} 
Note that the universality of this prefactor is connected to
the third law of thermodynamics -- a finite residual entropy 
per volume ($y_0=0$) would spoil this result.

It is rather difficult to measure thermal expansion inside a pressure
cell.  However, in many systems doping acts like ``chemical
pressure''. If doping $x$ and pressure $p$ can be quantitatively
related, $p-p_c=c (x-x_c)$, a measurement of $\Gamma$ for different
samples at ambient pressure can be used to check the prediction
(\ref{gr2}) quantitatively.
For generic tuning parameters, we need to substitute 
$(\partial r /\partial p)$ for $1/p_c$ in 
Eqs.\ (\ref{alpha-T},\ref{alpha-r}) and modify 
Eqs.\ (\ref{gr1},\ref{gr2}) accordingly. 

Similarly, for a QCP tuned by magnetic field [$r=(H-H_c)/H_c$] one
obtains for the magnetocaloric effect
\begin{eqnarray} 
\GE_{H,cr}(T\to 0,r)= -\frac{(\partial M/\partial T)_H}{c_{cr}}= 
- G_r  \frac{1}{H-H_c}. 
\label{grH} 
\end{eqnarray} 
Again, in the $T\to 0$ limit the prefactor (\ref{Cr}) is
universal.  The $T$-dependence of $\Gamma_{H,cr}$ at $r=0$ is also
given by (\ref{gr1}). 

It is interesting to compare the above with the case of a quantum
critical {\em line}, where the critical behavior is not restricted to
a single point but to a finite (pressure) interval.  Here, since only
marginal and irrelevant operators exist for $T=0$, $\Gamma_{cr}$ can
diverge at most logarithmically
\begin{eqnarray} 
\Gamma_{cr} \sim \pm \log T \,. 
\end{eqnarray} 
Conversely, if $\Gamma$ diverges algebraically for $T\to0$ a critical
line scenario can be excluded.

{\em Applicability of scaling:} The applicability of the scaling
results (\ref{scaling}--\ref{grH}) depends on a number of assumptions.
Most importantly, in an actual experiment not $\Gamma_{cr} =
\alpha_{cr}/c_{cr}$ but $\Gamma = \alpha/c$ is measured and sometimes
non-critical contributions can dominate (for an example see
below).  To verify Eqs.\ (\ref{gr1}) and (\ref{gr2}) in a situation where
$c_{cr}$ is subleading, one would have to subtract carefully the
non-critical contributions to the specific heat.

Generally, the scaling {\it Ansatz} (\ref{scaling}) holds only below
the upper critical dimension ($d+z<4$ within $\Phi^4$ theories).
At the upper critical dimension, logarithmic
corrections to scaling arise.  
Above
the upper critical dimension, the scaling argument can be spoiled by
the presence of so called ``dangerously irrelevant operators'': the
free energy is a singular function of irrelevant variables. Explicit
calculations (see below) for the case of an SDW transition
\cite{Hertz,Millis} show that on the paramagnetic side the irrelevant
operator
at most leads to logarithmic corrections.

A more subtle question is whether one of the basic assumptions
underlying the scaling approach (\ref{scaling}) holds: Is there a
single diverging time-scale close to the QCP? For example, in a nearly
magnetic metal the answer to this question is not obvious as there are
at least two types of low-energy degrees of freedom: magnetic
fluctuations and fermionic
quasiparticles\cite{belitz,roschPRB,Chubukov}.  This can indeed lead
to a breakdown of simple scaling relations as shown e.g. by Belitz
{\it et al.}  \cite{belitz}.
In the case of a {\em local} critical point  induced by a
(non-local) magnetic transition, as has been suggested by one of the
authors in \onlinecite{Si_Nature}, two scaling dimensions need
to be considered: one associated with the tuning of the 
long-wavelength fluctuations and the other 
with the tuning of the local fluctuations.

{\it SDW transitions:} We now turn to more specific calculations at
SDW quantum critical points, for two reasons.  First, they allow us to
address a number of questions concerning the scaling results: How do
corrections to scaling arise at the upper critical dimension? Are the
scaling results valid above this dimension? What happens, if the
prefactor $d-y_0 z$ in Eqs.  (\ref{gr2},\ref{grH}) vanishes?  Second,
our calculations are important for the purpose of assessing the
relevance of SDW QCPs to the magnetic quantum phase transitions in
heavy fermion compounds.

Our starting point is the Ginzburg--Landau--Wilson functional of Hertz
\cite{Hertz}:
\begin{eqnarray} 
S[\phi]&=& \sum _{ {\bf q}, i\omega _n} 
\left(\delta + q^2 + \frac {|\omega _n|}{ \Gamma _q}\right) 
|{\bf \phi}_{{\bf q}, i\omega _n}|^2+S^{(4)} \label{S}\,, \\ 
S^{(4)} &=& u \int _0^{\beta}d\tau \int d^d {\bf r}~ [{\bf \phi}({\bf r}, \tau)]^4 \,,
\nonumber 
\end{eqnarray} 
with $\Gamma _q=\Gamma _0 q^{z-2}$, where $z=2$ for an
antiferromagnetic SDW transition in a metal.  The $z=3$ theory may be
used to describe the critical endpoint of a metamagnetic first order
transition \cite{millis2} in $d=2,3$.  In the case of a ferromagnetic
QCP in $d=3$, the model (\ref{S}) with $z=3$ is only valid up to
logarithmic corrections and breaks down in $d=2$ \cite{belitz2}.  For
commensurate 2D magnetism coupled to 2D fermions there are additional
singularities in the fermion-collective-mode coupling\cite{Chubukov};
these singularities are absent when the fermions are taken to be
3D\cite{Kotliar}.  Following the renormalization group  scheme
adopted by Millis \cite{Millis,Zuelicke}
we have calculated the thermal expansion
and the Gr\"uneisen ratio
 for $d=2,3$ and $z=2,3$ on the non-magnetic side of
the phase diagram, $\delta \ge \delta_c$.  Details of the calculation
will be reported elsewhere.
  
\begin{table}[h] 
\begin{tabular}{ccccc} 
& $ d=2, z=3$ & $d=3, z=2$ & $ d=3, z=3$ & $d=2, z=2$  \\ 
\tableline 
&&\\ 
$\alpha_{cr}\sim$ & $T r^{-3/2}$ & $T r^{-1/2}$ & $T r^{-1}$ & $T r^{-1}$ \\ 
&&\\ 
$c_{cr}\sim$ & $T r^{-1/2}$ & $-T r^{1/2}$ & $T \log\frac{1}{r}$ & $T \log\frac{1}{r}$\\ 
&&\\ 
$\Gamma_{r,cr} = $ & $ (2 r)^{-1}$ & $- (2 r)^{-1}$ & $\left(r \log\frac{1}{r}\right)^{-1}$ & 
$\left(r \log\frac{1}{r}\right)^{-1}$ 
\end{tabular} 
\caption{\label{tab:FLregime} 
Results for SDW-QCPs 
in the Fermi liquid regime $r = \delta-\delta_c \gg T^{2/z}$. 
For a pressure tuned QCP, one obtains 
$\Gamma_{cr}=(dr/dp) \Gamma_{r,cr}/V_m$ using 
$r=(p-p_c)/p_c$, 
and $\Gamma_{H,cr}=(d r/dH) \Gamma_{r,cr}$ for 
$r=(H-H_c)/H_c$. Non-universal prefactors of $\alpha_{cr}$ and $c_{cr}$ 
are not shown. 
The prefactors of  $\Gamma_{cr}$ and 
$\Gamma_{H,cr}$ are (up to the logarithmic correction for $d=z$) 
universal. 
Note that for $d=3, z=2$ the specific heat is dominated by a  non-critical 
contribution $c_p\sim T$.} 
\end{table} 
  
\begin{table}[h] 
\begin{tabular}{ccccc} 
& $ d=2, z=3$ & $d=3, z=2$ & $ d=3, z=3$ & $d=2, z=2$  \\ 
\tableline 
&&\\ 
$\alpha_{cr} \sim$ & 
$\log \frac{1}{T}$ & $T^{1/2}$ & $T^{1/3}$ & 
$\log \log \frac{1}{T}$ \\ 
&&\\ 
$c_{cr} \sim$ & $T^{2/3}$ & $-T^{3/2}$ & $T \log \frac{1}{T}$ & $T \log \frac{1}{T}$\\ 
&&\\ 
$\Gamma_{r,cr} \sim$ & $T^{-2/3} \log \frac{1}{T} 
$ & $-T^{-1}$ & $\left(T^{2/3} \log\frac{1}{T}\right)^{-1}$ & 
$\frac{\log\log\frac{1}{T}}{T \log\frac{1}{T}}$ 
\end{tabular} 
\caption{\label{tab:QCregime} Results for SDW-QCPs 
in the quantum critical regime $r = \delta-\delta_c \ll T^{2/z}$ 
(cf. table \ref{tab:FLregime}).} 
\end{table} 
 
The results are summarized in Tables \ref{tab:FLregime} and 
\ref{tab:QCregime}. Up to logarithmic corrections the results obey the 
scaling forms (\ref{scaling}--\ref{grH}) with $\nu=1/2, y_0=1$. Note 
that for $d=z$ the pre\-factor in (\ref{scaling:thermalExpansion2}) and 
(\ref{gr2},\ref{grH}) vanishes. The $1/r$ dependence of $\alpha_{cr}$ 
for $d=z$ arises from a $T^2 \log 1/r$ correction to $F_{cr}$ not 
captured by scaling. 
For the quantum critical regime in $d=1/\nu=2$ the thermal expansion 
is logarithmic. The argument of the logarithm is a power of $T$ for 
$d+z > 4$ and is itself logarithmically dependent on $T$ for 
$d+z=4$; these features reflect the dangerously irrelevant or marginal nature of 
the quartic coupling $u$. 
 
In addition to the critical contributions, the measured quantities
also contain non-critical background components.  We list here the
full results for the purpose of 
comparisons with
experiments in heavy fermion compounds undergoing an antiferromagnetic
transition ($z=2$).  Consider first $d=3$.  At the QCP ($r=0$)
\begin{eqnarray} 
\alpha &=& a_1 T^{1/2} + a_2 T \,,
\label{d=3z=2QC} 
\end{eqnarray} 
where the $a_2$ term comes from the (fermionic) background
contribution. However, approaching the QCP in the Fermi-liquid regime
\begin{eqnarray} 
\alpha &=& ( { a_1  / {r^{1/2}}} + a_2 ) T \,.
\label{d=3z=2FL} 
\end{eqnarray} 
For $d=2$ and $z=2$, we have at the QCP ($r=0$)
\begin{eqnarray} 
\alpha &=& a_1 \log [ b \log \frac{T_0}{T} ] + a_2 T \,,
\label{d=2z=2QC} 
\end{eqnarray} 
and in the Fermi-liquid regime approaching the QCP: 
\begin{eqnarray} 
\alpha &=&  (a_1 / r + a_2) T \,.
\label{d=2z=2FL} 
\end{eqnarray} 

In two dimensions, the thermal expansion at $r=0$ diverges in the zero
temperature limit in sharp contrast to the textbook statement that
$\alpha(T \rightarrow 0) = 0 $.  Still,
it is straightforward to show that our results satisfy the
third law of thermodynamics.  
As $\alpha=
- (1/V) \partial S/\partial p$,
we can
write, for generic pressure,
\begin{eqnarray} 
S(p, T) = S(p_c, T) - \int _{p_c}^{p_*} \alpha V \, dp 
-    \int _{p_*}^{p} \alpha V \, dp \,,
\label{entropy} 
\end{eqnarray} 
where $p_*$ characterizes the crossover between the QC and FL regimes.
$S(p,T \to 0 ) \to 0$
due to a vanishing integration region [$(p^*-p_c) \propto T$] over
which $\alpha$ is divergent.

We now briefly discuss the experimental implications of our results.
Many heavy fermion compounds have long been known to show a
Gr\"uneisen ratio that increases to a very large value as temperature
is lowered\cite{Lacerda,Kambe}. 
Experiments are also becoming available in
the heavy fermion metals tuned to an antiferromagnetic quantum
critical point, making possible 
a systematic comparison with our theory\cite{Dresden}.
It is hoped that the present paper will
stimulate similar measurements in other kinds of (real and putative)
quantum critical materials.


In conclusion, we argue that the
Gr\"uneisen ratio
and the magnetocaloric effect
are divergent at any QCP. In addition, they
can be used to measure the 
scaling dimensions and to check the
very existence of a quantum critical point.

We would like to acknowledge helpful discussions with P. Gegenwart,
H.v.~L\"ohneysen, J. Mydosh, C. Pfleiderer, F. Steglich, and
J. D. Thompson. This work has been supported in part by NSF Grant
No.\ DMR-0090071, TCSAM, the Welch foundation (L.Z. and Q.S.)
and the Emmy-Noether program of the Deutsche Forschungsgemeinschaft
(M.G. and A.R.).


\begin{references}
\bibitem{gruen} The ``effective'' or ``thermodynamic'' Gr\"uneisen 
  parameter is usually defined as $\GE c_p/ (\kappa_T c_V)$, where $\kappa_T$
is the compressibility. We think that our $\GE$ is both experimentally and
theoretically the more convenient quantity.
Furthermore, 
neither $\kappa_T$ nor $c_p/c_V$ 
is expected to be 
singular at a magnetic QCP.
\end{references}
\end{document}